\documentclass[aps,prc,showpacs,showkeys,superscriptaddress,nofootinbib]{revtex4}
\usepackage{graphicx}

\newcommand{\be}{\begin{equation}}
\newcommand{\ee}{\end{equation}}
\newcommand{\bea}{\begin{eqnarray}}
\newcommand{\eea}{\end{eqnarray}}
\newcommand{\eq}[1]{Eq.~(\ref{#1})}
\newcommand{\boldtau}{\mbox{\boldmath $\tau$}}

\date{\today}
\begin{document}

\title{Renormalization of the leading-order chiral nucleon-nucleon interaction
and bulk properties of nuclear matter}
\author{R. Machleidt}
\email{machleid@uidaho.edu}
\affiliation{Department of Physics, University of Idaho, Moscow, Idaho 83844, USA}
\author{P. Liu}
\affiliation{Department of Physics, University of Idaho, Moscow, Idaho 83844, USA}
\author{D. R. Entem}
\email{entem@usal.es}
\affiliation{Grupo de F\'isica Nuclear, IUFFyM, Universidad de Salamanca, E-37008 Salamanca, Spain}
\author{E. Ruiz Arriola}
\email{earriola@ugr.es}
\affiliation{Departamento de F\'isica At\'omica, Molecular y  Nuclear, Universidad de Granada,
E-18071 Granada, Spain}

\begin{abstract}
We renormalize the two-nucleon interaction at leading order (LO) in
chiral perturbation theory using the scheme proposed by Nogga,
Timmermans, and van Kolck---also known as modified Weinberg
counting. With this interaction, we calculate the energy per nucleon
of symmetric nuclear matter in the Brueckner pair approximation and
obtain a converged, cutoff-independent result that shows saturation,
but also substantial underbinding. We find that the renormalized
LO interaction is characterized by an extraordinarily strong tensor
force (from one-pion exchange), which is the major cause for the lack
of binding. The huge tensor force also leads to the
unusually large wound integral of 40\% in nuclear matter, which
implies a very slow convergence of the hole-line or coupled-cluster expansion, 
rendering this interaction
impractical for many-body calculations.
In view of the unusual properties of the
renormalized LO interaction and in view of the poor convergence of the
nuclear many-body problem with this interaction, there is doubt if
this interaction and its predictions can serve as a reasonable and
efficient starting point that is improved by perturbative corrections.
\end{abstract}
\maketitle

\section{Introduction}
The problem of a proper derivation of nuclear forces is as old as
nuclear physics itself, namely, almost 80 years.  The modern view is
that, since the nuclear force is a manifestation of strong
interactions, any serious derivation has to start from quantum
chromodynamics (QCD).  However, the well-known problem with QCD is
that it is non-perturbative in the low-energy regime characteristic
for nuclear physics.  For many years this fact was perceived as the
great obstacle for a derivation of nuclear forces from
QCD---impossible to overcome except by lattice QCD.  The effective
field theory (EFT) concept has shown the way out of this dilemma.  One
has to realize that the scenario of low-energy QCD is characterized by
pions and nucleons interacting via a force governed by spontaneously
broken approximate chiral symmetry.  This chiral EFT allows for a
systematic low-momentum expansion known as chiral perturbation theory
(ChPT)~\cite{Wei79}.  Contributions are analyzed in terms of powers of
small momenta over the large scale: $(Q/\Lambda_\chi)^\nu$, where $Q$
is generic for a momentum (nucleon three-momentum or pion
four-momentum) or pion mass and $\Lambda_\chi \approx 1$ GeV is the
chiral symmetry breaking scale.  The early applications of ChPT
focused on systems like $\pi\pi$~\cite{GL84} and $\pi N$~\cite{GSS88},
where the Goldstone-boson character of the pion guarantees that the
expansion converges.  The past 15 years have also seen great progress
in applying ChPT to nuclear
forces~\cite{Wei90,Wei92,ORK94,Kol94,Kol99,KBW97,KGW98,EGM98,BK02,Kai01,EM02,EM02a,EM03,EGM05,ME05,Mac07,EHM08}.

However, there is a difference between the purely-pionic and the
one-nucleon sector, on the one hand, and two- and multi-nucleon
systems, on the other hand.  Nuclear physics is characterized by bound
states that are nonperturbative in nature. Weinberg
showed~\cite{Wei90} that the strong enhancement of the amplitude
arises from purely nucleonic intermediate states (``infrared
enhancement'').  He therefore suggested a two-step procedure: In step
one, ChPT and naive dimensional analysis is used to calculate a
``potential'' which consists of only irreducible diagrams and, in
step two, this potential is iterated to all orders by inserting it
into a Schr\" odinger or Lippmann-Schwinger (LS) equation to generate
the amplitude.

At leading order (LO), the potential consists of static one-pion
exchange (1PE) and two non-derivative contact terms.  At
next-to-leading order (NLO), multi-pion exchange starts which involves
divergent loop integrals that need to be regularized.  An elegant way
of doing this is dimensional regularization which (besides the main
nonpolynomial result) typically generates polynomial terms with
coefficients that are, in part, infinite or scale
dependent~\cite{KBW97}.  One reason why so-called contact terms are
introduced in the EFT is to absorb all infinities and scale
dependencies and make sure that the final result is finite and scale
independent.  This is the renormalization of the perturbatively
calculated $NN$ amplitude (which, by definition, is the ``$NN$
potential'').  It is very similar to what is done in the ChPT
calculations of $\pi\pi$ and $\pi N$ scattering, namely, a
renormalization order by order, which is the method of choice for any
EFT.  Thus, up to this point, the $NN$ calculation fully meets the
standards of an EFT and there are no problems.  The perturbative $NN$
amplitude can be used to make model independent predictions for
peripheral partial waves~\cite{KBW97,KGW98,EM02a}.

For calculations of the structure of nuclear few and many-body systems,
the lower partial waves are the most important ones. The fact that
in $S$ waves we have large scattering lengths and shallow (quasi)
bound states indicates that these waves need to be treated nonperturbatively.
Following Weinberg's prescription~\cite{Wei90}, this is accomplished by
inserting the potential $V$ into the LS equation:
\bea
 {T}({\vec p}~',{\vec p}) &=& {V}({\vec p}~',{\vec p}) \, 
+ \int \frac{d^3p''}{(2\pi)^3} \:
{V}({\vec p}~',{\vec p}~'')\:
\frac{M_N}
{{ p}^{2}-{p''}^{2}+i\epsilon}\:
{T}({\vec p}~'',{\vec p}) \,,
\label{eq_LS}
\eea
where $M_N$ denotes the nucleon mass.

In general, the integral in
the LS equation is divergent and needs to be regularized.
One way to achieve this is  by
multiplying $V$
with a regulator function, e.~g.,
\begin{equation}
{ V}(\vec{ p}~',{\vec p}) 
\longmapsto
{ V}(\vec{ p}~',{\vec p})
\;\mbox{\boldmath $e$}^{-(p'/\Lambda)^{2n}}
\;\mbox{\boldmath $e$}^{-(p/\Lambda)^{2n}}
\label{eq_regulator} \,.
\end{equation}
Typical choices for the cutoff parameter $\Lambda$ that
appears in the regulator are 
$\Lambda \approx 0.5 \mbox{ GeV} \ll \Lambda_\chi \approx 1$ GeV~\cite{EM03,EGM05}.

It is pretty obvious that results for the $T$-matrix may
depend sensitively on the regulator and its cutoff parameter.
This is acceptable if one wishes to build models.
For example, the meson models of the past~\cite{MHE87,Mac89}
always depended sensitively on the choices for the
cutoff parameters which were, in fact,
welcome fit-parameters for achieving a good reproduction of the $NN$ data.
However, the EFT approach wishes to be fundamental
in nature and not just another model.

In field theories, divergent integrals are not uncommon and methods have
been developed for how to deal with them.
One regulates the integrals and then removes the dependence
on the regularization parameters (scales, cutoffs)
by renormalization. In the end, the theory and its
predictions do not depend on cutoffs
or renormalization scales.

So-called renormalizable quantum field theories, like QED, have
essentially one set of prescriptions that takes care of
renormalization through all orders.  In contrast, EFTs are
renormalized order by order, in which case the number of adjustable
parameters increases.

As discussed, the renormalization of {\it perturbative} EFT
calculations is not a problem. {\it The problem is nonperturbative
  renormalization.}  This problem typically occurs in {\it nuclear}
EFT because nuclear physics is characterized by bound states which are
nonperturbative in nature.

Weinberg's implicit assumption was that the counterterms introduced to
renormalize the perturbatively calculated potential, based upon naive
dimensional analysis (``Weinberg counting''), are also sufficient to
renormalize the nonperturbative resummation of the potential in the LS
equation.  Unfortunately, it has turned out that this assumption is
not quite correct, as pointed out by Kaplan, Savage, and Wise
(KSW)~\cite{KSW96}, and others. The criticism of the Weinberg counting
scheme resulted in a flurry of publications on the renormalization of
the nonperturbative $NN$
problem~\cite{FMS00,PBC98,FTT99,Bea02,VA05-1,NTK05,VA05-2,VA07,Bir06,Bir09,EM06,VA08,Ent08,YEP07,LK08,BKV08,Val08,EG09,Val09}.
The literature is too comprehensive to discuss all contributions in
detail.  Let us just mention some of the work that has particular
relevance to our present paper.

If the potential $V$ consists of contact terms only (a.k.a.\ pion-less
theory), then the nonperturbative summation, Eq.~(\ref{eq_LS}), can be
performed analytically, which makes it easier to deal with the
renormalization issue.  However, when pion exchange is included, then
Eq.~(\ref{eq_LS}) can be solved only numerically and the
renormalization problem is less transparent.  Perturbative ladder
diagrams of arbitrarily high order, where the rungs of the ladder
represent a potential made up from irreducible pion exchange, suggest
that an infinite number of counterterms is needed to achieve cutoff
independence for all the terms of increasing order generated by the
iterations.  For that reason, Kaplan, Savage, and Wise
(KSW)~\cite{KSW96} proposed to sum the contact interaction to all
orders (analytically) and to add pion exchange perturbatively up to
the given order. Unfortunately, it turned out that the order by order
convergence of 1PE is poor in the $^3S_1$-$^3D_1$ state~\cite{FMS00}.
The failure was triggered by the $1/r^3$ singularity of the 1PE tensor
force when iterated to second order. Therefore, KSW counting is no
longer taken into consideration (see, however, Ref.~\cite{BKV08}).  
A balanced discussion of possible
solutions can be found in Ref.~\cite{Bea02}.

Some researchers decided to take a second look at Weinberg's original
proposal.  A systematic investigation and re-analysis of Weinberg
counting in leading order has been conducted by Nogga, Timmermans, and
van Kolck~\cite{NTK05} in momentum space, and by Pav\'on Valderrama and
one of the present authors (ERA) at LO and higher orders in
configuration space~\cite{VA05-1,VA05-2,VA07}. A comprehensive
discussion of both approaches and their equivalence can be found
in Refs.~\cite{Ent08,Val08}.  The LO $NN$ potential consists of 1PE plus two
nonderivative contact terms that contribute only in $S$ waves. By
numerical calculations\footnote{For the purposes of the present paper,
  we conduct the discussion in momentum space. We note however, that
  all information on the necessary number of counterterms can be
  determined {\it a priori} and analytically by inspecting the
  potential in configuration space at short
  distances~\cite{VA05-1,VA05-2,VA07,Ent08}.}, Nogga {\it et al.}  find
that the given counterterms renormalize the $S$ waves, i.e., the
naively expected infinite number of counterterms is not needed. This
means that Weinberg power counting does actually work in $S$ waves at
LO (ignoring the $m_\pi$ dependence of the contact interaction
discussed in Refs.~\cite{KSW96,Bea02}).  However, there are problems
with a particular class of higher partial waves, namely those in which
the tensor force from 1PE is attractive. The first few cases of this
kind of low angular momentum are $^3P_0$, $^3P_2$, and $^3D_2$. The
leading order (nonderivative) counterterms do not contribute in $P$
and higher waves, which is the reason for the problem. But the second
order contact potential provides counterterms for $P$
waves. Therefore, the promotion of, particularly, the $^3P_0$ and
$^3P_2$ contacts from NLO to LO would fix the problem in $P$ waves. To
take care of the $^3D_2$ problem, a fourth order contact needs to be
promoted to LO.  In this way, one arrives at a scheme of `modified
Weinberg counting'~\cite{NTK05} for the leading order two-nucleon
interaction.

Once cutoff independence of the on-shell $NN$ $T$-matrix (and $NN$
phase shifts and observables) has been achieved, it is of interest to
know if cutoff independent results are also obtained when this
interaction is applied in nuclear few- and many-body systems.  Nogga
{\it et al.}~\cite{NTK05} investigated the lightest such system,
namely, the three-nucleon bound state, and found cutoff independence
of the triton binding energy.  It is the purpose of this note to
conduct a similar investigation in heavier nuclear systems. Since
finite nuclei are difficult to calculate, we choose nuclear matter
(infinitely many nucleons). We will show that the renormalized LO
two-nucleon interaction leads to converged results for the energy per
nucleon in nuclear matter.

Section~II briefly describes and repeats the LO renormalization procedure 
with modified Weinberg counting introduced in Ref.~\cite{NTK05}.
In Sec.~III, we present the novel point of this paper,
namely, the calculation of the energy per nucleon in symmetric nuclear matter
as a function of density and, in Sec.~IV, we compare our results with the work
by other authors. Conclusions are drawn in Sec.~V.

\section{Renormalizing the $NN$ potential in leading order}

\begin{figure}[t]
\vspace*{-1.5cm}
\hspace*{-1.5cm}
\scalebox{0.5}{\includegraphics{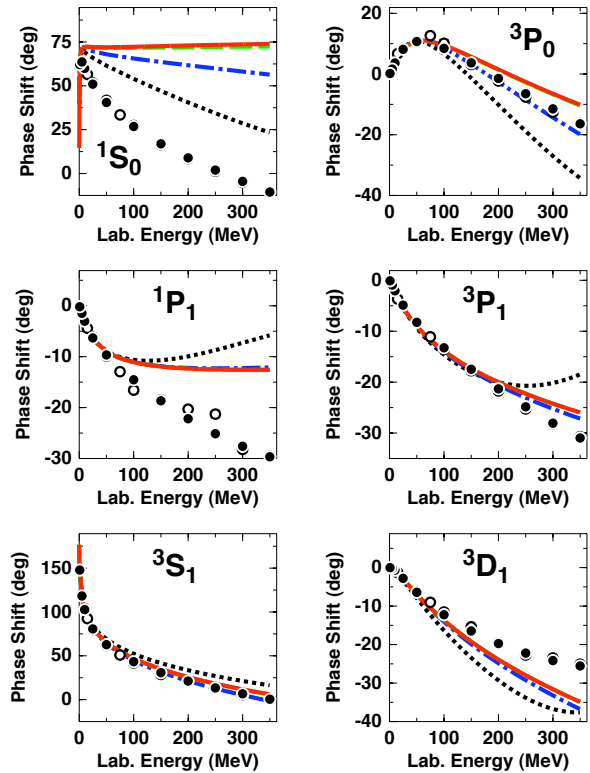}}
\vspace*{-2.0cm}
\caption{(color online). Phase shifts and mixing parameters of 
neutron-proton scattering for total
angular momentum $J\leq 2$ and $T_{\rm lab}\leq 350$ MeV. The curves display the LO predictions
for cutoff parameter $\Lambda=0.5$~GeV (black dotted), 1~GeV (blue dash-dotted), 5~GeV
(green dashed), and 10~GeV (red solid).
 Note that the dashed and the solid curves are, in general,
indistinguishable on the scale of the figure. 
The solid dots and open circles represent the results from the Nijmegen
multi-energy $np$ phase shift analysis~\protect\cite{Sto93} 
and the VPI/GWU
single-energy $np$ analysis SM99~\protect\cite{SM99}, respectively.}
\end{figure}

\begin{figure}[t]
\vspace*{-1.5cm}
\hspace*{-1.5cm}
\scalebox{0.5}{\includegraphics{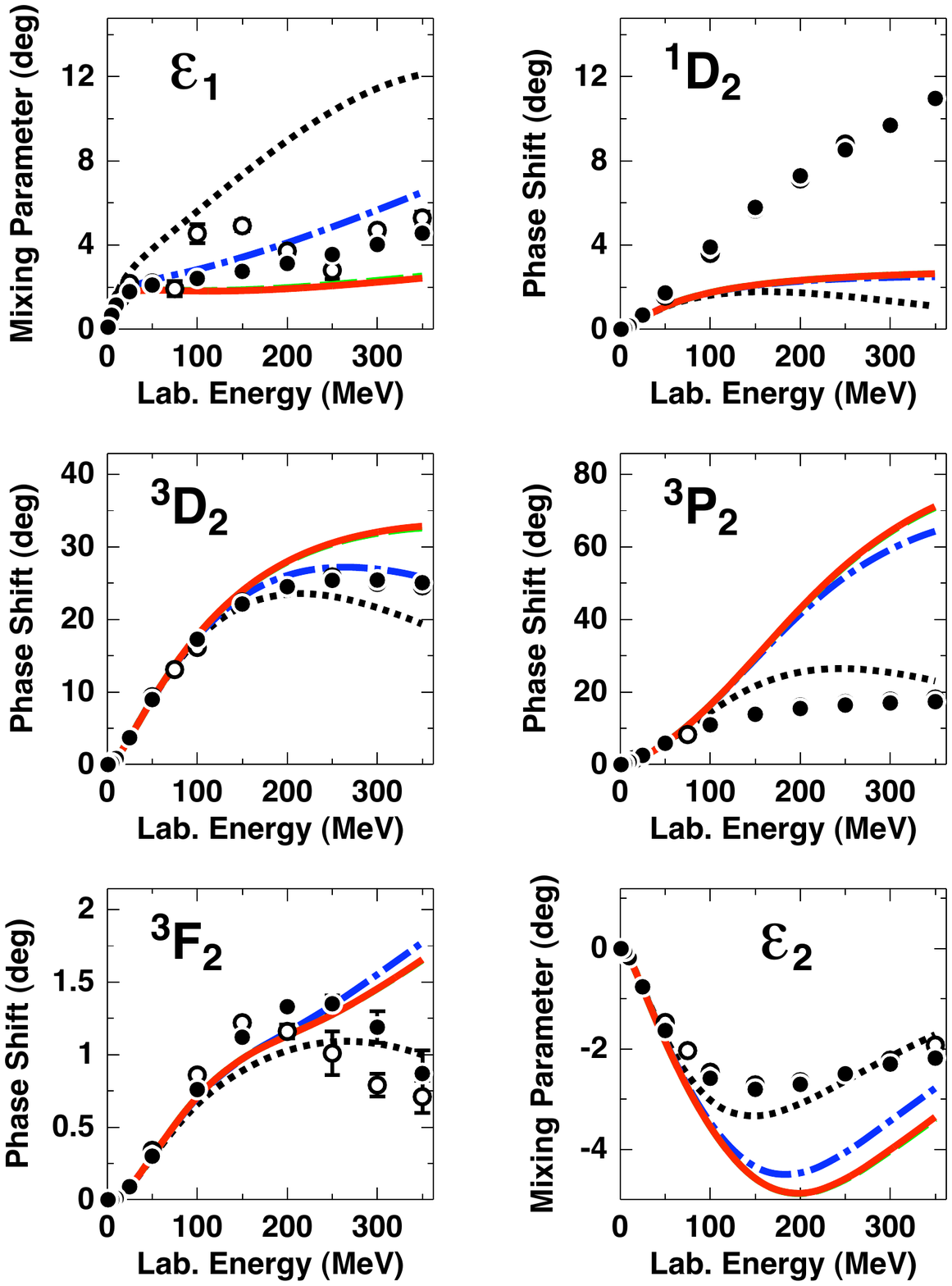}}

\vspace*{-1.5cm}

FIG.~1, continued.
\end{figure}

In naive dimensional analysis (``Weinberg counting''), the
order by order expansion of the chiral $NN$ potential is given as:
\bea
V_{\rm LO} & = & 
V_{\rm ct}^{(0)} + 
V_{1\pi}^{(0)} 
\label{eq_VLO}
\\
V_{\rm NLO} & = & V_{\rm LO} +
V_{\rm ct}^{(2)} + 
V_{1\pi}^{(2)} +
V_{2\pi}^{(2)} 
\label{eq_VNLO}
\\
V_{\rm NNLO} & = & V_{\rm NLO} +
V_{1\pi}^{(3)} + 
V_{2\pi}^{(3)} 
\label{eq_VNNLO}
\\
V_{{\rm N}^3{\rm LO}} & = & V_{\rm NNLO} +
V_{\rm ct}^{(4)} +
V_{1\pi}^{(4)} +  
V_{2\pi}^{(4)} +
V_{3\pi}^{(4)} 
\label{eq_VN3LO}
\eea where the superscript denotes the order $\nu$ of the low-momentum
expansion.  LO stands for leading order, NLO for next-to-leading
order, etc..  Contact potentials carry the subscript ``ct'' and
pion-exchange potentials can be identified by an obvious subscript.

The charge-independent one-pion exchange (1PE) potential reads
\begin{equation}
V_{1\pi} ({\vec p}~', \vec p) = - 
\frac{g_A^2}{4f_\pi^2}
\: 
\boldtau_1 \cdot \boldtau_2 
\:
\frac{
\vec \sigma_1 \cdot \vec q \,\, \vec \sigma_2 \cdot \vec q}
{q^2 + m_\pi^2} 
\,,
\label{eq_1PEci}
\end{equation}
where ${\vec p}~'$ and $\vec p$ designate the final and initial
nucleon momenta in the center-of-mass system (CMS) and $\vec q \equiv
{\vec p}~' - \vec p$ is the momentum transfer; $\vec \sigma_{1,2}$ and
$\boldtau_{1,2}$ are the spin and isospin operators of nucleon 1 and
2; $g_A$, $f_\pi$, and $m_\pi$ denote axial-vector coupling constant,
the pion decay constant, and the pion mass, respectively. We use
$f_\pi=92.4$ MeV and $g_A=1.29$ to correct for the Goldberger-Treiman
discrepancy.  Since higher order corrections contribute only to mass
and coupling constant renormalizations and since, on shell, there are
no relativistic corrections, the on-shell 1PE has the form
\eq{eq_1PEci} in all orders.

Here, we will specifically calculate LO neutron-proton ($np$) scattering 
and take charge-dependence (isospin violation) into account.
Thus, the 1PE potential reads
\begin{equation}
V_{1\pi}^{(np)} ({\vec p}~', \vec p) 
= -V_{1\pi} (m_{\pi^0}) + (-1)^{I+1}\, 2\, V_{1\pi} (m_{\pi^\pm})
\,,
\label{eq_1penp}
\end{equation}
where $I$ denotes the isospin of the two-nucleon system and
\begin{equation}
V_{1\pi} (m_\pi) \equiv - \,
\frac{g_A^2}{4f_\pi^2} \,
\frac{
\vec \sigma_1 \cdot \vec q \,\, \vec \sigma_2 \cdot \vec q}
{q^2 + m_\pi^2} 
\,.
\end{equation}
We use $m_{\pi^0}=134.9766$ MeV and
 $m_{\pi^\pm}=139.5702$ MeV~\cite{PDG}.
In the LS equation, \eq{eq_LS}, we apply
\bea
M_N  &=&  \frac{2M_pM_n}{M_p+M_n} = 938.9182 \mbox{ MeV,}
\\
p^2 & = & \frac{M_p^2 T_{\rm lab} (T_{\rm lab} + 2M_n)}
               {(M_p + M_n)^2 + 2T_{\rm lab} M_p}  
\,,
\eea
where $M_p=938.2720$ MeV and $M_n=939.5653$ MeV
are the proton and neutron masses~\cite{PDG}, respectively, and 
$T_{\rm lab}$ 
is the kinetic energy of the incident neutron 
in the laboratory system (``Lab.\ Energy'').
The relationship between $p^2$ and
$T_{\rm lab}$ 
is based upon relativistic kinematics.

Besides the 1PE potential, \eq{eq_1penp}, the EFT includes contact terms
which represent short-range interactions that cannot be resolved at the
low-energy scale. Furthermore, the contacts are needed for renormalization.
Stating the contact potentials in partial-wave decomposition,
we have one zero-order ($\nu=0$) contact in each $S$ wave:
\bea
V^{(0)}_{\rm ct} (^1S_0) &=& 
\widetilde{C}_{^1S_0} 
\,, 
\label{eq_ct_1s0}
\\
V^{(0)}_{\rm ct} (^3S_1) &=& 
\widetilde{C}_{^3S_1} 
\,. 
\eea
Up to this point, we are still applying Weinberg counting. However, as discussed in the
Introduction, higher partial waves in which the pion's tensor force is attractive
need counter terms to achieve cutoff independence---which leads us to
modified Weinberg counting. To be specific, two $P$ waves receive counter terms
of second order,
\bea
V^{(2)}_{\rm ct} (^3P_0) &=& 
C_{^3P_0} 
\,\, p'p \,, \\
V^{(2)}_{\rm ct} (^3P_2) &=& 
C_{^3P_2} 
\,\, p'p \,, 
\eea
and one $D$ wave needs a fourth-order counter term,
\bea
V^{(4)}_{\rm ct} (^3D_2) &=& 
D_{^3D_2} 
\,\, {p'}^2 p^2 \,. 
\label{eq_ct_3d2}
\eea

For the solution of the LS equation, \eq{eq_LS}, 
a regulator function is necessary, for which we choose the one given in \eq{eq_regulator}
with $n=2$. The regulator depends on the cutoff mass $\Lambda$, which we vary over
a wide range from 0.5 GeV to 10 GeV. In $S$-waves, we readjust the contact parameter
for each choice of $\Lambda$ such that the empirical scattering lengths 
($a_s=-23.748$ fm for $^1S_0$ and $a_t=5.4170$ fm for $^3S_1$)
are reproduced.
In those $P$ and $D$ waves which carry a contact in modified Weinberg counting, 
the contact parameter is used to 
fit---for the various choices of $\Lambda$---the 
empirical phase shift at 50 MeV as given in Ref.~\cite{Sto93}.
In all cases, we then calculate the phase shifts for all energies below 350 MeV.

The resulting phase shifts and mixing parameters for total angular
momentum $J\leq 2$ are shown in Fig.~1. The curves refer to $\Lambda =
0.5$ GeV (dotted), 1 GeV (dash-dotted), 5 GeV (dashed), and 10 GeV
(solid).  The curves for $\Lambda=$ 5 GeV and 10 GeV are, in general,
indistinguishable on the scale of the figure, which demonstrates that
cutoff independence (nonperturbative renormalization) has been
achieved. It is known from the work of Nogga {\it et al.}~\cite{NTK05}
in momentum space\footnote{Again, the analysis is easier in
  configuration space~\cite{VA05-1,VA05-2,VA07}; see, in particular,
  Table II and the thorough convergence analysis
  of phase shifts with total angular momentum $J \le 5$ in Ref.~\cite{VA07}.}  that this
is possible.  Our results represent an independent confirmation.

\begin{table}
\caption{Partial-wave contact parameters as a function of the cut-off $\Lambda$ for the 
leading-order $NN$ potential in modified Weinberg counting.
The parameters are defined in Eqs.~(\ref{eq_ct_1s0})-(\ref{eq_ct_3d2}).}
\begin{tabular}{ccccc}
\hline
\hline
Partial-wave contact & \multicolumn{4}{c}{--- Cutoff parameter $\Lambda$ in units of GeV ---}\\
parameter & 0.5 & 1.0 & 5.0 & 10.0 \\
\hline
$\widetilde{C}_{^1S_0}$ 
($10^4$ GeV$^{-2}$)
&--0.109966&--0.087189&--0.06739623&--0.064460345
\\
$\widetilde{C}_{^3S_1}$ 
($10^4$ GeV$^{-2}$)
&--0.076005&1.349900&--0.02692560&0.021786000
\\
           $ C_{^3P_0} $
($10^4$ GeV$^{-4}$)
&0.840321&--0.1722517&0.001856514&0.000384981
\\
          $  C_{^3P_2} $
($10^4$ GeV$^{-4}$)
&--0.2316105&--0.0700665&--0.00251447&0.001251038
\\
           $ D_{^3D_2} $
($10^4$ GeV$^{-6}$)
&--0.3347880&0.3899800&--0.00020581&--0.00001055
\\
\hline
\hline
\end{tabular}
\label{tab_ct}
\end{table}

\section{Nuclear matter}

As discussed in the Introduction, once cutoff independence has been
achieved for the two-nucleon system, a good question to ask is if
cutoff-independent predictions are also obtained in the nuclear
many-body problem when this renormalized $NN$ potential is applied.
Nogga {\it et al.}~\cite{NTK05} addressed this question for the
three-nucleon system where they confirmed the cutoff independence of
the triton binding energy at LO.  We wish to turn to heavier nuclear
systems and choose nuclear matter as the representative sample.

By definition, nuclear matter refers to an infinite uniform system of nucleons
interacting via the strong force without electromagnetic interactions.
This hypothetical system is believed to approximate conditions in the
interior of heavy nuclei. We shall assume equal neutron and proton densities,
that is, we will consider symmetric nuclear matter. This many-body system is 
characterized by its energy per nucleon as a function of the particle density.

We will use the well-established Brueckner-Bethe-Goldstone method (in
short: Brueckner theory)~\cite{Mac89,Day67,Bet71,HT70} to calculate
the nuclear matter energy.  In this theory, a central role is played
by the Brueckner $G$-matrix which is a solution of the Bethe-Goldstone
integral equation
\begin{equation}
G(w)=V-V\frac{Q}{H_0-w}G(w) \,,
\label{eq_G}
\end{equation}
where $w$ denotes the starting energy, $H_0$ the unperturbed Hamiltonian, and
the Pauli operator $Q$ projects onto unoccupied states. In the pair approximation,
the energy per nucleon is given by
\begin{equation}
\frac{E}{A}=\frac{1}{A} \sum_{m\leq k_F} \langle m|t|m\rangle +
\frac{1}{2A} \sum_{m,n\leq k_F} \langle mn|G(w)|mn-nm\rangle
\end{equation}
where $A$ denotes the number of nucleons, $t$ the kinetic energy operator,
and $k_F$ the Fermi momentum, which is related to the density $\rho$ of symmetric nuclear
matter by
\begin{equation}
\rho=\frac{2}{3\pi^2}k_F^3 \,.
\end{equation}
The starting energy is chosen on-shell, i.e.
\begin{equation}
w=e(m)+e(n)
\end{equation}
with single-particle energy
\begin{equation}
e(m)=t(m)+U(m)
\end{equation}
and single-particle potential
\begin{equation}
U(m)=\left\{ \begin{array}{ll}
             \sum_{n\leq k_F} \langle mn|G(w)|mn-nm\rangle, & m\leq k_F \\
                                    0                     , & m > k_F \\
             \end{array}
     \right.
\label{eq_gap}
\end{equation}
also known as the ``gap'' choice for the single-particle potential,
since a gap will obviously occur at the Fermi surface.
The calculations are conducted in partial-wave decomposition
and the Brueckner integral-equation is solved by matrix inversion,
see Ref.~\cite{HT70} for details.

\begin{figure}[t]
\vspace*{-1.5cm}
\scalebox{0.35}{\includegraphics{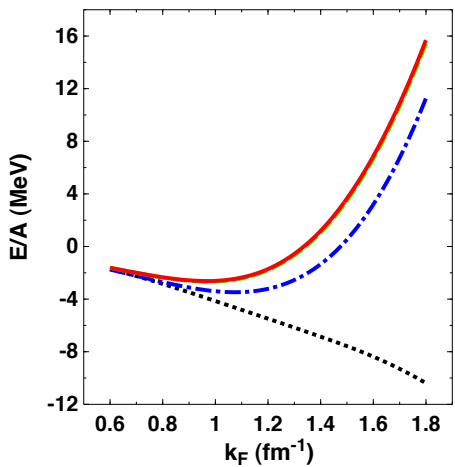}}
\vspace*{-2.0cm}
\caption{(color online). Energy per nucleon, $E/A$, in symmetric  nuclear matter
as a function of the Fermi momentum $k_F$ applying the LO $NN$ potentials
with the various cutoffs used in the phase-shift calculations of Fig.~1.
The curve patterns represent the same cutoffs as in Fig.~1.}
\end{figure}

The Bethe-Goldstone method was originally devised to handle the short
distance hard core in nuclear systems. In chiral EFT, an expansion
both in $1/f_\pi$ and $1/M_N$ is carried out. For short distances $r
\ll 1/m_\pi$ pion mass effects can be neglected and on purely
dimensional grounds an inverse power short distance singularity should
be expected: to LO $V(r) \sim 1/(f_\pi^2 r^3)$, to NLO $V(r) \sim
1/(f_\pi^4 r^5)$, to NNLO, $V(r) \sim 1/(f_\pi^4 M_N r^6) $ etc. It is
natural to ask whether the $G-$matrix result converges in the limit
$\Lambda \to \infty$. In the appendix we show that this is indeed the
case for the gap choice Eq.~(\ref{eq_gap})~\cite{HT70}. The proof
rests on the finiteness of the off-shell $K$-matrix~\cite{Ent09}. This
gives some confidence on the stability of the numerics for increasing
cut-off values.

In Fig.~2, we display our results for the energy per nucleon in
symmetric nuclear matter as a function of density (measured by the
Fermi momentum $k_F$) applying the LO $NN$ potentials with the various
cutoffs used in the phase-shift calculations of Sec.~II.  The same
curve patterns in Figs.~1 and 2 indicate the same cutoffs.  The
nuclear matter curves for $\Lambda=5$ GeV (green dashed) and
$\Lambda=10$ GeV (red solid) cannot be distinguished on the scale of
Fig.~2, demonstrating that cutoff independence of the predictions is
achieved; or in other words, the red solid curve represents the
renormalized result which, as expected, is convergent. This curve shows
saturation at a Fermi momentum $k_F\approx 1.0$ fm$^{-1}$ and an
energy per nucleon $E/A=-2.6$ MeV.

Based upon various pieces of circumstantial evidence, it is generally
believed that the ``empirical'' saturation properties of symmetric
nuclear matter are $k_F=1.35\pm 0.05$ fm$^{-1}$ and $E/A=-16\pm 1$
MeV~\cite{Mac89}.  Thus, our renormalized LO result shows considerable
underbinding.  In Ref.~\cite{NTK05}, a triton energy of $-3.6$ MeV was
found for the converged LO result which also deviates considerably
from the empirical value of $-8.5$ MeV.

{\it The chief reason for this lack of attraction is the fact that the
  tensor force of the renormalized LO interaction is unusually
  strong,} as we will explain now.

A simple indicator for the strength of the tensor force component
contained in a given $NN$ potential is the predicted $D$-state
probability of the deuteron, $P_D$, because the transition from $S$ to
$D$ states can only proceed via the tensor force.  For the LO
interaction at $\Lambda=5$ GeV and 10 GeV, the $P_D$ comes out to be
7.2\% (it's converged). Conventional potentials predict $P_D$
typically lower, namely, between 4 and 6\%; for example, the
AV18~\cite{WSS95}, CD-Bonn~\cite{Mac01}, and N3LO~\cite{EM03}
potentials predict 5.76\%, 4.85\%, and 4.51\%,
respectively. Historically, the largest $P_D$ ever predicted by a
``realistic'' $NN$ potential was 7.0\% by the Hamada-Johnston
potential~\cite{HJ62} of 1962.

\begin{table}
\caption{Partial-wave contributions for $S$ and $P$ waves and total
  contributions to symmetric nuclear matter at a density equivalent to
  a Fermi momentum $k_F=1.35$ fm$^{-1}$. Unless denoted otherwise;
  numbers without parentheses or brackets state contributions to the
  potential energy as obtained from using the Brueckner $G$ matrix;
  numbers in parentheses are corresponding results obtained in Born
  approximation, i.e., for $G=V$; finally, figures in brackets state
  contributions to the wound integral.}
\begin{tabular}{ccccc}
\hline
\hline
Partial wave & LO$^a$ & N3LO$^b$ & CD-Bonn\cite{Mac01} & AV18\cite{WSS95}  \\
\hline
\hline
               & --13.42 & --16.74 & --16.76 & --16.07 \\
$^1S_0$ & (--5.98) &(--14.73) & (--12.64) & (--2.77) \\
               & [0.056] & [0.008] & [0.005] & [0.017] \\
\hline
               & --13.54 & --19.42 & --18.96 & --17.10 \\
$^3S_1$ & (+10.65) & (--12.51) & (--8.63) & (+5.99) \\
               & [0.090, 0.112]$^c$ & [0.017, 0.017]$^c$ & [0.004, 0.034]$^c$ & [0.015, 0.053]$^c$ \\
\hline
               &3.24 & 3.90 & 3.91 & 3.88 \\
$^1P_1$ & (3.27) & (4.06) & (4.24) & (4.23) \\
               & [0.000] & [0.001] & [0.002] & [0.001] \\
\hline
               & --1.01 & --3.14 & --3.08 & --3.15 \\
$^3P_0$ & (--6.30) & (--3.03) & (--2.41) & (--2.48) \\
               & [0.109] & [0.001] & [0.002] & [0.002] \\
\hline
               & 10.17 & 9.68 & 9.81 & 9.74 \\
$^3P_1$ & (11.09) & (10.29) & (11.74) & (12.08) \\
               & [0.004] & [0.003] & [0.006] & [0.007] \\
\hline
               & --5.37 & --7.27 & --7.05 & --6.96 \\
$^3P_2$ & (--1.49) & (--6.96) & (--6.25) & (--6.09) \\
               & [0.015, 0.015]$^d$ & [0.001, 0.001]$^d$ & [0.003, 0.001]$^d$ & [0.002, 0.002]$^d$ \\
\hline
\hline
Total potential & --22.44 & --37.02 & --36.35 & --33.96 \\
energy            & (+9.07) & (--26.72) & (--17.92) & (+6.96) \\
\hline
Kinetic energy & 22.67 & 22.67 & 22.67 & 22.67\\
\hline
Total    & +0.23 & --14.35 & --13.67 & --11.29 \\
energy & (+31.74) & (--4.05) & (+4.76) & (+29.63) \\
\hline
Total wound & [0.405] & [0.050] & [0.058] & [0.101] \\
\hline
\hline
\end{tabular}

\begin{flushleft}
\footnotesize $^a$The renormalized LO $NN$ potential of this work with
$\Lambda = 10$ GeV.\\ $^b$Quantitative N$^3$LO $NN$ potential
regularized by a Gaussian with cutoff parameter $\Lambda = 0.5$
GeV~\cite{EM03}.\\ $^c$The $^3S_1$-$^3S_1$ and the $^3S_1$-$^3D_1$
contributions to the wound integral are given.\\ $^d$The
$^3P_2$-$^3P_2$ and the $^3P_2$-$^3F_2$ contributions to the wound
integral are given.
\end{flushleft}
\label{tab_nm}
\end{table}

In nuclear matter, the so-called wound integral $\kappa$ is known to
depend sensitively on the strength of the tensor
force~\cite{HT70,Mac89}. The wound integral is defined as
\begin{equation}
\kappa = \rho \int |\phi-\psi|^2 d\tau \,,
\end{equation}
where $\phi$ denotes the uncorrelated two-nucleon wave function and $\psi$ the correlated one,
which are related by
\begin{equation}
G\phi = V\psi \,,
\end{equation}
implying
\begin{equation}
\psi=\phi-\frac{Q}{H_0-w}G\phi \,.
\end{equation}
The physical significance of the wound integral is that it measures
the probability for exiting two nucleons to states above the Fermi
surface. This probability is large for ``hard'' and strong-tensor
force potentials.  According to arguments conveyed by
Brandow~\cite{Bra67}, a $n$-hole line diagram is proportional to
$\kappa^{n-1}$ and, hence, the convergence of the hole-line expansion
depends on the size of $\kappa$, with large $\kappa$ suggesting slow
convergence.

As shown in the bottom row of Table ~II, the renormalized LO
interaction produces a total $\kappa$ of 40.5\%, while the
corresponding numbers are 10.1\%, 5.8\%, and 5.0\% for AV18, CD-Bonn,
and N3LO. The Hamada-Johnston potential generated a total $\kappa$ of
21.1\%. The partial-wave contributions to $\kappa$ listed in Table~II
(numbers in square brackets) confirm that the strong tensor force of
the LO interaction is the main reason for the extraordinarily high
$\kappa$. The $^3S_1$-$^3D_1$ transition, which depends entirely on
the tensor force, contributes 11.2\% to the LO $\kappa$, while it's
5.3\%, 3.4\%, and 1.7\% for AV18, CD-Bonn, and N3LO, respectively. In
Fig.~3, we show the $^3S_1$-$^3D_1$ transition potential of the LO
interaction with $\Lambda=10$ GeV and of conventional potentials
revealing dramatic differences, particularly, for high momenta.  An
unusual difference occurs also in the $^3P_0$ state where LO generates
a contribution to $\kappa$ of 10.9\%, whereas conventional potentials
have at most 0.2\%. The tensor operator is known to have a large
matrix element in the $^3P_0$ state. The $^3P_0$ potentials are
included in Fig.~3.

The fact that a strong tensor force (and a large $\kappa$) leads to
less binding energy in nuclear matter and finite nuclei can be
understood as follows~\cite{Mac89}.  For the purpose of discussion,
let's approximate the Brueckner $G$ by
\begin{equation}
G(w)\approx V_C-V_T\frac{Q}{H_0-w}V_T \,,
\label{eq_Gapprox}
\end{equation}
where $V_C$ denotes the central force and $V_T=v_TS_{12}$ the tensor
force component of a given $NN$ potential (with $S_{12}$ the usual
tensor operator).  Note now that all quantitative nuclear potentials
are fit to the same $NN$ data and, thus, produce essentially the same
on-shell $T$-matrix or, equivalently, the same on-shell $K$-matrix,
which in the above approximation is given by
\begin{equation}
K(w_f)\approx V_C-{\cal P}V_T\frac{1}{t-w_f}V_T \,,
\label{eq_Kapprox}
\end{equation}
where $w_f$ is the free (purely kinetic) starting energy and $\cal P$
denotes the principal value.

\begin{figure}[t]
\vspace*{-0.8cm}
\scalebox{0.25}{\includegraphics{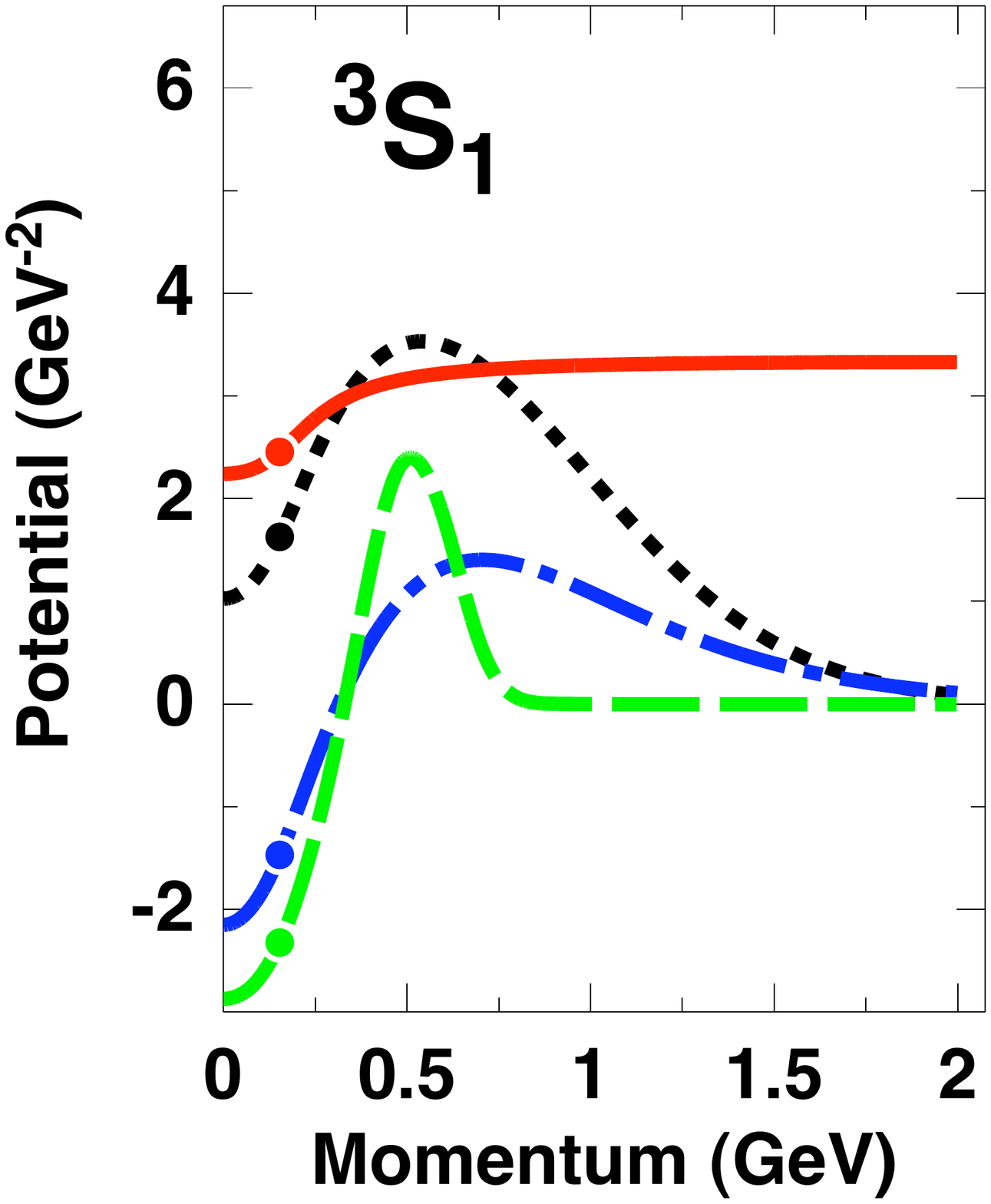}}
\scalebox{0.25}{\includegraphics{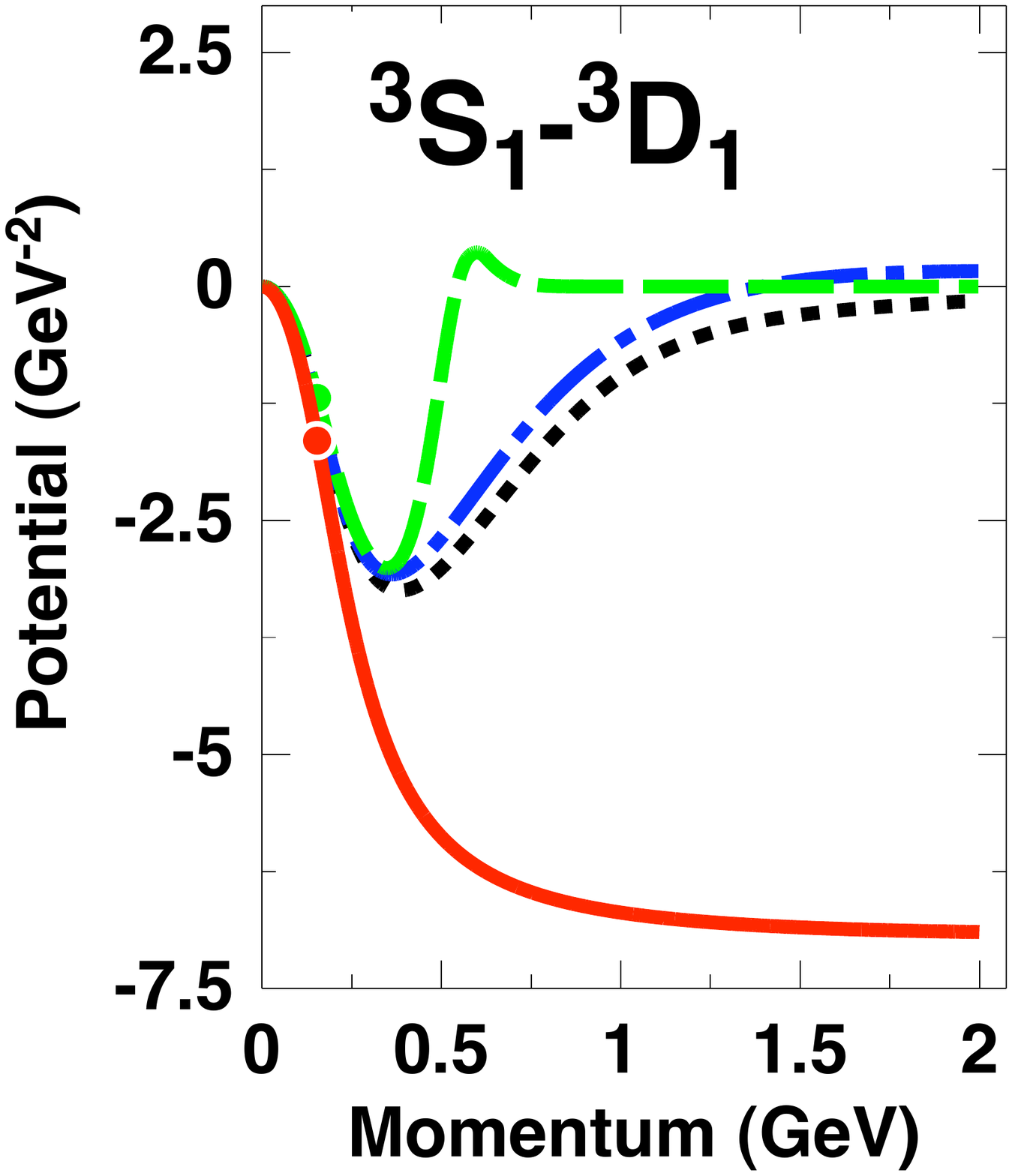}}
\scalebox{0.25}{\includegraphics{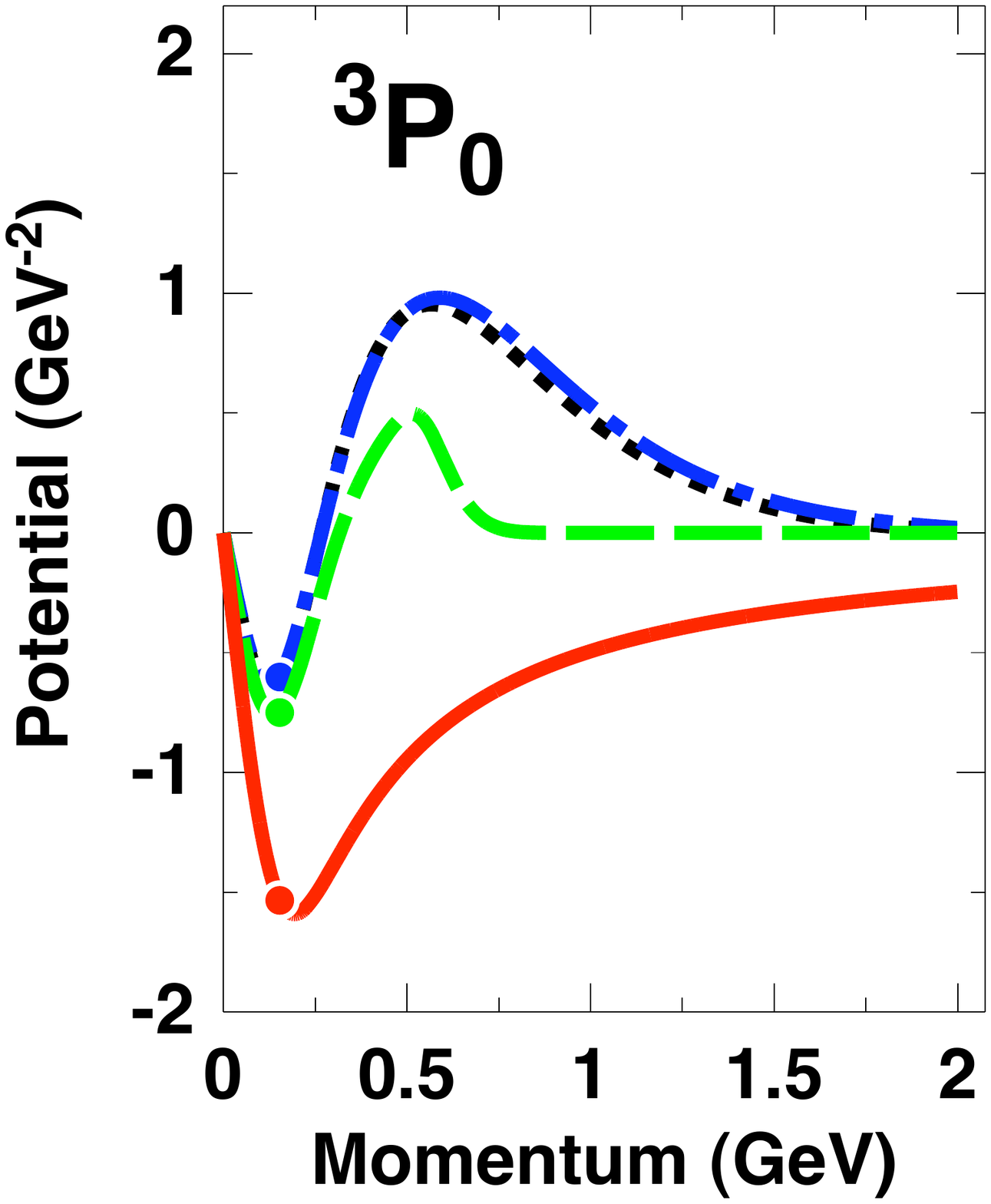}}
\vspace*{-1.0cm}
\caption{(color online). Half off-shell $NN$ potentials, $V/(2\pi)^3$,
  for partial-wave states as denoted.  The potentials displayed are
  Argonne V18 (black dotted), CD-Bonn (blue dash-dotted), N3LO (green
  dashed), and the renormalized chiral LO potential of Sec.~II for
  $\Lambda=10$ GeV (red solid).  The on-shell point at $p=153$ MeV/c
  (equivalent to $T_{\rm lab}=50$ MeV) is marked by a solid dot.}
\end{figure}

A potential with a strong $V_T$ (implying a large, attractive second order
  tensor term) will have a less attractive central force $V_C$ to
arrive at the same on-shell $K$-matrix as compared to a potential with
a weak tensor force. Now, when we enter nuclear matter and calculate
the $G$-matrix Eq.~(\ref{eq_Gapprox}), the Pauli operator $Q$ [which
  is absent in the free-space Eq.~(\ref{eq_Kapprox})] and a larger
energy denominator (due to the single particle potential in the
many-body environment) reduces the magnitude of the second term of the
$G$-matrix equation.  These two medium effects are know as the Pauli
and dispersion effects.  The larger the attractive second order tensor term in
Eq.~(\ref{eq_Gapprox}), the larger the reduction of the attraction through the medium
effects.  Therefore, potentials which produce large integral terms in
the $G$-matrix equation will predict less attraction in the many-body
system.  When the central force is very strong (``hard'' potential),
the above mechanism applies also to the iterations of the central
term. This happens obviously in the $^1S_0$ state where no tensor force is
involved, but never-the-less a large $\kappa$ occurs for the LO
interaction.  This is also part of the reason why the $^3S_1$-$^3S_1$
contribution to the wound is large for LO, namely 9.0\% (cf. the very
hard LO central force seen in the $^3S_1$ frame of Fig.~3).

An idea of the size of the integral term in the Brueckner equation,
Eq.~(\ref{eq_G}), is also obtained by comparing the Born approximation
(i.e., $G=V$) with the full $G$.  We therefore provide in Table~II
also the Born approximation results (numbers in parentheses) for the
various partial-wave contributions.

As explained in length in Ref.~\cite{Bra88}, arguments similar to the
above also apply to Faddeev calculations of the three-particle
energy. Thus, the substantial underbinding of the triton found in
Ref.~\cite{NTK05} is most likely also related to the huge tensor force
of the renormalized LO interaction.

We note that the discussion of the severely reduced attraction in
nuclear matter due to ``hard'' central potentials and large tensor
forces applies, of course, only to a calculation conducted in the
Brueckner pair approximation. The huge wound integral suggests that
there will be large three, four, and higher hole-line contributions
which may provide additional binding. However, the evaluation of
multi-hole line contributions is extremely involved and cumbersome.
Similar arguments apply when the coupled-cluster expansion is used to
deal with the nuclear many-body problem~\cite{Hag08}.  Thus, an
extraordinarily strong tensor force makes it very difficult to obtain
converged results in the many-body system, which is one reason why a
large tensor-force potential is inconvenient, to say the least.

The best studied phenomenology of nuclear forces is the
one-boson-exchange model.  This model includes a $\rho$ meson, which
produces a tensor force of opposite sign as compared to the pion
(cf. Fig.~3.7 of Ref.~\cite{Mac89}).  Careful studies have shown that
the reduction of the pion's tensor force at short range by the
$\rho$-meson is crucial to arrive at a realistic strength for the
nuclear tensor force~\cite{BM94}.

In ChPT theory, contributions from heavy mesons, like the $\rho$, are
too short-ranged to be dissolved, but instead contact terms are added
to the theory. The set of contacts which appears at NLO ($\sim Q^2$)
includes a tensor term that may be perceived as simulating $\rho$
exchange. Therefore, there is a chance that, at NLO or higher order, the problem of
the extraordinarily large tensor force encountered at LO will be
resolved.

\section{Comparison with other works and the broader perspective}

Our results clearly show that the saturation mechanism in nuclear
matter is compatible with a non-perturbative renormalization of the
venerable 1PE potential.  However, the binding energy turns out to be
rather small and the approach looks discouraging from a
coupled-cluster expansion point of view. Let us therefore analyze our
results in the light of other chiral approaches to nuclear matter,
based in spirit on the EFT concept, where mainly perturbative schemes
but also low cut-offs have been employed. We will also provide some
perspective on future work.

If the cut-off parameter takes sufficiently small values,
perturbation theory becomes applicable, since in such a case high
momentum components are suppressed. For the smallest cut-off value
presented in Fig.~2, $\Lambda=0.5$ GeV, we see a clear attraction which
is strongly dependent, in fact linearly, on the Fermi
momentum. Actually, chiral symmetry based approaches for nuclear
matter have also been pursued in Ref.~\cite{KFW02} in a purely
perturbative scheme. The main source of attraction stems from
once iterated 1PE, is proportional to the density, $\sim k_F^3$, and
depends linearly also on the cut-off parameter, which needs to be fine-tuned 
to a value $\Lambda=0.4-0.5$ GeV to achieve saturation.
The origin of the divergence is related to vacuum amplitudes which to
n-th order with the 1PE potential yield a contribution to the
$T$-matrix, Eq.~(\ref{eq_LS}), scaling by naive power counting as $\sim
\Lambda^{n-1} $ (pion mass neglected). The present calculation
contains iterated 1PE {\it with} additional counterterms to all orders
for {\it any} value of the cut-off parameter and, as we see, it does
not exhibit this very strong $k_F^3$ dependent attraction.  Thus, in
Fig.~2 only some residual cut-off effect is displayed {\it after} low
energy $NN$ physics has been fixed.  Clearly, the separately large
perturbative contributions which scale with positive powers of
$\Lambda$ would not converge without inclusion of counterterms to all
orders.
 
A NLO $G$-matrix calculation with a finite cut-off of $\Lambda=500$
MeV and counterterms was also undertaken in Ref.~\cite{SA08} and
saturation was found. As noted in \cite{VA05-2}, there would be a
fundamental problem of removing the cut-off at that order, since the
deuteron becomes unbound due to the strong $1/r^5$ repulsive
interaction in the triplet channels. More recently, a perturbative
approach has also been proposed where a power-counting scheme is
introduced with extremely low cut-off values $\Lambda \sim m_\pi$ for
which saturation is achieved~\cite{OM09,OM09bis}.

Finally, $V_{\rm low-k}$ approaches represent a coarse graining of the
interaction in the physically accessible $NN$ elastic region for CM
momenta $k \le \Lambda \approx 400$  MeV with the result that {\it all}
high precision potentials fitting data with $\chi^2/DOF\approx 1$ including
the 1PE tail collapse into a unique $V_{\rm low-k}$ potential. It has
been suggested~\cite{Bo05} that low-momentum interactions in general
and chiral N$^3$LO interactions (having their own cut-off
$\Lambda \approx 500-700$  MeV~\cite{EM03,EGM05}), in particular,
can be treated perturbatively in nuclear matter calculations, as the
corresponding Weinberg eigenvalues lie inside the unit circle. Given
the universality of the $V_{\rm low-k}$ approach and the fact that all
interactions contain 1PE while successfully describing the phase shifts
in vacuum, there seems to be no fingerprint left of chiral dynamics
from the two-body sector. When three-body chiral forces determined
from few-body data are included, realistic saturation properties with
more controlled uncertainties are obtained~\cite{Bo09}. A trading
between two and three body forces is observed and the role played by
3N forces as the $V_{\rm low-k}$ cut-off is increased from
$\Lambda=400$  MeV to $\Lambda=560$ MeV becomes less
important although they still produce saturation at realistic densities.

Clearly, {\it explicit} chiral dynamics is enhanced for larger cut-off
values but also the theoretical difficulties increase. The spirit of
the original proposal of Weinberg's was that the potential could be
perturbatively defined according to a prescribed counting. This is
clearly possible at large impact parameters where neither the strength
of the nuclear force nor our lack of knowledge of its short distance
components prevent us from using a fuzzy but sensibly large momentum cut-off value.  In
a sense, the $G$-matrix approach is close in spirit to this idea, where
the strength of the nucleon force in the nuclear medium is
characterized by an effective interaction. As we have shown, the
LO 1PE interaction yields cut-off independent results, is entirely
parameterized by vacuum properties, and has the nice feature of
saturation. The problem is to define {\it what} is meant by NLO, and in
nuclear matter a scheme tightly close to the standard perturbative EFT
idea is expected to face the same problems already found in the vacuum
sector and described in the Introduction.

Several possible and perturbatively motivated schemes for the
renormalization of the chiral $NN$ interaction suggested already in
Ref.~\cite{NTK05} include considering higher pion exchanges as
perturbations around the LO calculation using distorted waves. This
proposal was analyzed in Refs.~\cite{VA05-2,Val09}, where it was shown
to be feasible and cut-off independent, but with little gain from a
practical viewpoint: many more counter terms were needed and a worse
description was obtained in the $^1S_0$ channel and the
deuteron. Actually, non-perturbative calculations with singular
potentials behave non-analytically in the coupling constant,
i.e. $1/f_\pi^{\alpha}$ with non-integer $\alpha$~\cite{VA05-2}. A
RG-based program for pion-full theories was advanced in
Ref.~\cite{Bir06}, with further detailed results provided in
Ref.~\cite{Bir09}.
This RG analysis yields one counter term in each $S$-wave of order $Q^{-1}$, which
must be iterated, and one
in each $^3P_J$ and $^3D_J$ wave of order $Q^{-1/2}$, which may be iterated.
This is very similar to the power counting suggested by Nogga {\it et al.}~\cite{NTK05}.
Subleading counterterms occur in $^1S_0$ at $Q^0$ and in $^3S_1$ at $Q^{1/2}$, 
while the subleading $^3P_J$ and $^3D_J$ terms are of order $Q^{3/2}$.
Subleading and higher terms ought to be treated perturbatively. 
Finally each $S$ wave receives an additional term at order $Q^2$.
The number of terms at $Q^3$ in the RG scheme is larger than in Weinberg
counting at NNLO; however, except for the subleading
$D$-wave interactions, it is essentially the same as Weinberg counting at N$^3$LO ($Q^4$).

 Toy models where a $1/r^2$ singular potential is
perturbed by $1/r^4$ interactions provide useful insight, but the
consequences for more realistic cases have not yet been worked
out~\cite{LK08}.  A follow-up KSW scheme has been pursued in
Ref.~\cite{BKV08}, where the short distance singularity is tamed by
the introduction of a Pauli-Villars like pion-mass of about the
$\rho$-meson mass. An important prerequisite for specific calculations
is the perturbative renormalizability described in
Ref.~\cite{Val09}. However, as we discussed in Ref.~\cite{Ent08} the
huge change needed from the simple LO $^1S_0$ phase to the real
observed one does not suggest any sort of small effect in the large
momentum region as non-perturbatively renormalized calculations
suggest. As already mentioned, this partial wave provides an important
contribution to the energy in nuclear matter.

Turning to non-perturbative schemes, it was noted in Ref.~\cite{Ent09}
that off-shell properties of renormalized chiral potentials do not
look much different from more conventional ones as the singularities
are effectively removed.  Furthermore, we know that the inclusion of
$\Delta$ degrees of freedom in chiral potentials~\cite{KGW98} lead to
a pattern of better convergence, where e.g. the NLO-$\Delta$ and
NNLO-$\Delta$ deuteron does exist~\cite{VA09}, with an acceptable
phenomenological success and where much larger and naively more
natural momentum cut-off values display a better convergence (see also
the second Ref.~\cite{EGM98} where $\Lambda \sim 1{\rm GeV}$ is
taken). Actually, this is a case where the discussion on
renormalizability becomes pointless as the renormalized and the
natural sized cut-offs are not too far apart, since finite cut-off
effects are less important the more singular the
potential~\cite{Ent08}. Indeed, at order $\nu $ in the chiral
counting the potential scales as $ 1/( \Lambda_\chi^{\nu+2} r^{3+\nu})$ and
then the finite cut-off correction, $\delta_\Lambda (k)$,  to the renormalized phase shifts, $\delta_\infty (k)$, behaves as $\delta_\infty (k) -
\delta_\Lambda (k) ={\cal O}(\Lambda^{-1/2-\nu/2})$~\cite{Ent08}.
Moreover, unlike the $\Delta$-less renormalized scheme, the deuteron
$D$-state probability becomes $P_D \sim 5.8 \%$, a comparable value to
conventional and phenomenologically successful potentials. According
to our discussion, the corresponding wound integral would be small
enough as to guarantee a good converging pattern of few body
correlations in the nuclear many body problem. Thus, a $G$-matrix
approach applied to chiral interactions including $\Delta$-isobar
degrees of freedom has some chance of furnishing the multiple
theoretical requirements of renormalizability, power counting for the
potential (a la Weinberg), and presumably convergence in the nuclear
medium. At present it is unclear whether such a scheme will be
phenomenologically successful as the potentials do not contain the
important spin-orbit contributions and it remains to be seen if those
can be represented by appropriate counterterms.

\section{Conclusions}

In this work, we have renormalized the two-nucleon interaction at
leading order (LO) in chiral perturbation theory using the scheme
proposed by Nogga, Timmermans, and van Kolck~\cite{NTK05}---also known
as modified Weinberg counting.  With this interaction, we have
calculated the energy of symmetric nuclear matter in the Brueckner
pair approximation.  We find that the energy per nucleon as a function
of nuclear matter density converges to a cutoff-independent (i.e.,
renormalized) result and shows saturation. The predicted value for the
energy per nucleon at saturation shows considerable underbinding,
which is in line with the converged LO prediction for the triton
binding energy of Ref.~\cite{NTK05}.  We demonstrate that the LO
interaction contains an unusually strong tensor force (from
pion-exchange), which is the main reason for the lack of binding in
few- and many-body systems.  In fact, the tensor force is stronger
than that of any $NN$ potential ever constructed in the 50-year
history of realistic nuclear forces.

The huge tensor force of the renormalized LO interaction leads to the
unusually large wound integral of 40\% in nuclear matter, which
implies a very slow convergence of the hole-line expansion and,
similarly, the coupled-cluster expansion, rendering this interaction
impractical for many-body calculations.

It is well-known from the meson-theory of nuclear forces that the
tensor force produced by the pion needs to be damped at short range
(or high momenta). In conventional models, this is achieved through
$\pi NN$ form factors and contributions from heavy-meson exchange
(particularly, $\rho$ exchange). ChPT, which does not include heavy
mesons, provides a contact term of tensor structure at NLO. Thus, a
more realistic tensor force may be expected at higher orders.

Several possible schemes for the renormalization of the chiral $NN$
interaction that have recently been given serious thought are designed to
renormalize the LO interaction non-perturbatively (as done in the
present work) and to add higher order corrections in perturbation
theory.  However, in view of the unusual properties of the
renormalized LO interaction and in view of the poor convergence of the
nuclear many-body problem with this interaction, there is doubt if
this interaction and its predictions can serve as a reasonable and
efficient starting point that is improved by perturbative corrections.
To make the interaction more suitable for many-body calculations, one may consider
to apply renormalization group methods~\cite{BFP07} to construct a soft two-nucleon force 
(2NF) plus
a three-nucleon force (3NF) equivalent to the original LO interaction. Then there will
be no problem with the application of this ``$V_{\rm low-k}$'' 2NF in the many-body system.
However, it may turn out that the 3NF (needed for proper equivalence with
the original LO interaction) is so strong that the issue is just shifted towards a problem
with the
convergence of the 3NF contribution.

\begin{appendix}

\section{Finiteness of $G$-matrix for singular interactions}

In this appendix, we show the finiteness of the $G$-matrix for singular
interactions such as the 1PE potential for the gap choice
Eq.~(\ref{eq_gap})~\cite{HT70}, which implies that the potential energy is
suppressed as compared to the kinetic energy in the large momentum
region. The idea underlying the proof is that short distance physics
much below the healing distance~\cite{Day67} does not depend on the
Fermi momentum. It is convenient to start with the extension of the $G$-matrix 
to any arbitrary energy $z$,  
\begin{eqnarray}
G(z) = V + V \frac{Q}{z-H_0} G(z)  \,,
\end{eqnarray}
and re-write this equation as 
\begin{eqnarray}
G(z)^{-1} = V^{-1} - \frac{Q}{z-H_0}  \,.
\end{eqnarray}
The $G$-matrix used in the main text, Eq.~(\ref{eq_G}) corresponds to
taking $z=w=e(m)+e(n)$. We now introduce the extended $K$-matrix defined as
\begin{eqnarray}
K(z) = V + V {\cal P} \frac{1}{z-t} K(z) \,,
\end{eqnarray}
where ${\cal P}$ denotes the principal value and $t$ is the kinetic
energy operator. The $K$-matrix used in the main text,
Eq.~(\ref{eq_Kapprox}), corresponds to using $z=w_f=t(m)+t(n)$. Below
threshold, the principal value prescription can be removed as the pole
is never hit. This equation can likewise be written as
\begin{eqnarray}
K(z)^{-1} = V^{-1} - {\cal P} \frac{1}{z-t}  \,.
\end{eqnarray}
In Ref.~\cite{Ent09} the finiteness of the $K$-matrix {\it off-shell}
for short distance singular interactions was established solely on the
basis of on-shell renormalization conditions. This means that if $K$
is finite on shell then the off-shell extension remains finite as
well. Note that the renormalization conditions are basically
equivalent to a {\it fixed} energy, e.g. zero energy. Thus, $K$ will
remain finite {\it also} below threshold. Taking $z=w$ and subtracting
$K(w)^{-1}$ from $G(w)^{-1}$ we have (principal value where ever
necessary)
\begin{eqnarray}
G(w) = K(w) + K(w) \left[ \frac{Q-1}{w-H_0} -\frac1{w-t} +
  \frac1{w-H_0} \right] G(w)
\end{eqnarray}
In this equation, the regulator may be effectively removed from the
$K$-matrix at the operator level even if the energy $w \neq w_f$ does
not correspond to the scattering energy ~\cite{Ent09}. If the
interaction is attractive the single particle potential $U(n)<0$ and
$w < w_f$ and the (finite cutoff) $K(w)$ involves states below
threshold. As we see, the first integral only involves states below
the Fermi surface and is thus bound and cut-off independent provided
$k_F \ll \Lambda$. The only remaining piece where the cut-off enters
explicitly is in the last too terms involving
$(w-H_0)^{-1}-(w-t)^{-1}$, which for the gap choice, see
Eq.~(\ref{eq_gap})~\cite{HT70}, vanishes for states above the Fermi
surface, making the integral convergent as well.  Thus, the finiteness
of the $G$-matrix is reduced to the finiteness of the $K$-matrix {\it
  off-shell}.

\end{appendix}

\section*{Acknowledgements}
This research was supported in part by the U.S. Department of Energy
under Grant No.~DE-FG02-03ER41270 (R.M.); the Ministerio de Ciencia y
Tecnolog\'\i a under Contracts No. FPA2007-65748, by Junta de Castilla
y Le\'on under Contract No.  SA-106A07 and GR12 (D.R.E.); by the European
Community-Research Infrastructure Integrating Activity ``Study of
Strongly Interacting Matter'' (HadronPhysics2 Grant No. 227431), the
Spanish Ingenio-Consolider 2010 Program CPAN (CSD2007-00042), the
Spanish DGI and FEDER funds with Grant No. FIS2008-01143/FIS, Junta de
Andaluc{\'\i}a Grants No.  FQM225-05, EU Integrated Infrastructure
Initiative Hadron Physics Project contract no. RII3-CT-2004-506078 (E.R.A.).

\bibliographystyle{unsrt}

\end{document}